Super-resolution discrete-Fourier-transform spectroscopy using precisely periodic radiation beyond time window size limitation

Takeshi Yasui,[1,2,3] Yuki Iyonaga,[1] Yi-Da Hsieh,[2,3] Yoshiyuki Sakaguchi,[1] Francis Hindle,[4] Shuko Yokoyama,[1,5] Tsutomu Araki,[1] and Mamoru Hashimoto[1]

[1]Graduate School of Engineering Science, Osaka University, 1-3 Machikaneyama, Toyonaka, Osaka 560-8531, Japan

[2]Institute of Technology and Science, Tokushima University, 2-1 Minami-Josanjima, Tokushima 770-8506, Japan

[3]JST, ERATO, MINOSHIMA Intelligent Optical Synthesizer Project, 2-1 Minami-Josanjima, Tokushima 770-8506, Japan

[4]Laboratoire de Physico-Chimie de l'Atmosphère, Université du Littoral Côte d'Opale, 189A Av. Maurice Schumann, Dunkerque 59140, France

[5]Micro Optics Co., Ltd, 2-20, Oe-Nakayama, Nishikyo, Kyoto 610-1104, Japan

Abstract

Fourier transform spectroscopy (FTS) has been widely used in a variety of fields in research, industry, and medicine due to its high signal-to-noise ratio, simultaneous acquisition of signals in a broad spectrum, and versatility for different radiation sources. Further improvement of the spectroscopic performance will widen its scope of applications. Here, we demonstrate improved spectral resolution by overcoming the time window limitation using discrete Fourier transform spectroscopy (dFTS) with precisely periodic pulsed terahertz (THz) radiation. Since infinitesimal resolution can be achieved at periodically discrete frequencies when the time window size is exactly matched to the repetition period $T$, a combination of THz-dFTS with a spectral interleaving technique achieves a spectral resolution only limited by the spectral interleaving interval. Linewidths narrower than $1/(50T)$ are fully resolved allowing the attribution of rotational-transition absorption lines of low-pressure molecular gases within a 1.25 MHz band. The proposed method represents a powerful tool to improve spectrometer performance and accelerate the practical use of various types of FTS.

# 1. Introduction

Fourier transform spectroscopy (FTS) is a spectroscopic technique where spectra are obtained by measuring a temporal waveform or interferogram of electromagnetic radiation, or other types of radiation, and calculating its Fourier transform (FT). FTS possesses inherent advantages over conventional dispersive spectrometers, such as a high signal-to-noise ratio (SNR), simultaneous acquisition of signals in a broad spectrum, and versatility for different radiation sources. Therefore, FTS has been widely used in various fields, such as FT infrared (FT-IR) spectroscopy [1], terahertz (THz) time-domain spectroscopy (THz-TDS) [2], FT electrical/optical spectral analysis, FT nuclear magnetic resonance (FT-NMR) [3], FT mass spectroscopy (FT-MS) [4], and so on.

When the temporal waveform of a phenomenon is measured, the spectral resolution is simply determined by the inverse of the measurement time window size during which the temporal waveform is observed [1-4]. Therefore, as the time window is increased, the spectral resolution is enhanced. However, when the majority of the signal components are temporally localized, excessive extension of the window size increases the noise contribution as well as the acquisition time. Furthermore, in the case of optical FTS, the travel range of a translation stage used for time-delay scanning practically limits the spectral resolution.

When the phenomenon repeats, it is generally accepted that the achievable spectral resolution is limited to its repetition frequency because the maximum window size is restricted to a single repetition period to avoid the coexistence of multiple signals. However, if the repeating phenomenon is observed using precisely periodic radiation, the time window may be expanded beyond a single period without accumulation of timing errors. Recently, dual comb spectroscopy (DCS) has emerged in the ultraviolet, visible, infrared, and even THz regions [5-12]. DCS acquires the temporal waveforms of repeating phenomena with a time window extending over many repetition periods and by FT achieves a spectral resolution finer than the repetition frequency. The spectral resolution is still limited by the time window, along with massive volume of datapoints required for rapid sampling over a long window. Such a huge number of data results in stringent restrictions for the data acquisition and FT calculation.

If the spectral resolution in FTS can be improved beyond the repetition frequency without the need to acquire the huge number of datapoints, the scope of application of FTS will be further extended. In this article, we demonstrate a significant spectral resolution imptovement over the time window limitation by using discrete Fourier transform spectroscopy (dFTS) in THz region. We show that this is only possible when the time window size is exactly matched to the repetition period of precisely periodic THz pulse.

## 2. Results

### 2.1 Principle of operation

First we consider the measured temporal waveform *h(t)* of a phenomenon and its FT spectrum *H(f)* given as

$$H(f) = \int_{-\infty}^{\infty} h(t)\exp(-2\pi ift)dt \qquad (1).$$

This equation indicates that a spectral component *H(f)* is obtained by multiplying *h(t)* by a frequency signal exp(-2πift) and then integrating the product for an infinite integration period. This process is illustrated in Fig. 1(a), where cos2πft is shown as the real part of exp(-2πift). Although the achieved spectral resolution is infinitesimal in Eq. (1), the practical resolution is limited by the achievable finite integration period due to SNR, the acquisition time, and/or the stage travel range.

Next we consider the case where *h(t)* is repeated by precisely periodic pulsed radiation with a repetition period of *T*, and the relaxation time of *h(t)* is longer than *T*. A series of signals *h(t)* temporally overlap, each subsequent events being delayed by integer multiple of *T* [see Fig. 1(b)]. When the time series data is acquired with a finite time window size $\tau$, the observed time window includes signal contributions from multiple periods, for example signals (A), (B), (C), and (D) as depicted in Fig. 1(c). To obtain the FT spectrum of these, the signals (A), (B), (C), and (D) are summed and multiplied by exp(-2πift) before being integrated over $\tau$ as shown in Fig. 1(d). Here, if *T* is sufficiently stable and is exactly matched to $\tau$, signals (A), (B), (C), and (D) can be temporally connected to form a single, temporally continuous signal alongside exp(-2πift) [see Fig. 1(e)]. Despite of the finite time window size (=$\tau$ = *T*), this connection is equivalent to acquiring the temporal waveform of *h(t)* without limitation of the time window size. Because of the smooth connection of exp(-2πift), an infinitesimal spectral resolution can be achieved at discrete frequencies

$$f_n = n/T \qquad (n = 0, 1, \cdots, N-1), \qquad (2)$$

where $n$ is the order of the spectral data points, and $N$ is the total number of sampling points in the temporal waveform [see the top of Fig. 1(f)]. The spectral data at the discrete frequencies are calculated simply by taking the discrete Fourier transform (dFT) of the signals in Fig. 1(c).

Although these spectral data points provide infinitesimal spectral resolution, their discrete spectral distribution limits the spectral sampling interval to *1/T*. To harness the improved spectral resolution for broadband spectroscopy, we must interleave additional marks of scale into the frequency gap between data points, namely, spectral interleaving [13-15]. If incremental sweeping of *T* is repeated to interleave the additional marks ($f_n'$, $f_n''$, $\cdots$) [see the midlles of Fig. 1(f)], and all of the resulting dFT spectra are overlaid, the frequency gaps can be filled in [see the bottom of Fig. 1(f)], allowing us to obtain a more densely distributed, discrete spectrum which has great potential for broadband high resolution spectroscopy. Since this procedure is equivalent to obtaining the dFT spectral components while tuning *T* with a fixed *N* in Eq. (2), the spectral sampling density becomes higher due to the increase in the number of data points. Furthermore, if *T* is phase-locked to a microwave frequency standard while maintaining $\tau = T$, the frequency interval between the spectral data points is universally constant, and hence the absolute frequency of the spectrum is secured to the frequency standard.

2.2 Experimental setup

To observe a repeating phenomenon precisely by using a stabilized, mode-locked THz pulse train and acquiring its temporal waveform with $\tau = T$, we applied the asynchronous optical sampling (ASOPS) method [16-19]. Since the ASOPS method enables us to expand the ps time scale of THz transient signals up to the μs scale, the resulting slowed-down temporal waveform can be directly and precisely measured by a data acquisition board without the need for mechanical time-delay scanning. This non-mechanical nature of the time-delay scanning enables us to correctly acquire the temporal waveform where $\tau$ is accurately matched to $T$ ensuring a smooth temporal connection of the signal contributions.

Figure 2 is a schematic diagram of the THz-dFTS setup, which contains 1.5-μm dual mode-locked Er-doped fibre lasers ($f_{rep1}$ = 250,000,000 Hz, $f_{rep2}$ = 250,000,050 Hz, and $f_{offset}$ = 50 Hz) and a THz optical setup for low-pressure gas spectroscopy. The temporal waveform of the precisely periodic THz signal was acquired at a sampling interval of 100 fs for a time window size of 4 ns (= $T$), corresponding to 40,000 sampling data, at a scan rate of 50 Hz. After signal accumulation, a THz discrete amplitude spectrum with a spectral sampling interval of 250 MHz was obtained by taking its dFT. Furthermore, we repeated a similar procedure at different $T$ by changing both $f_{rep1}$ and $f_{rep2}$, and interleaved additional marks of scale in the dFT spectrum [15].

2.3 Spectroscopy of low-pressure water vapour

To investigate the importance of consistency between $\tau$ and $T$ in the temporal connection of Fig. 1(e), we performed low-pressure spectroscopy of water vapour. Water vapour shows sharp absorption lines in the THz region due to rotational transitions of the asymmetric top molecule. Here, the rotational transition $1_{10} \leftarrow 1_{01}$ at 0.557 THz was measured. To satisfy the condition that the phenomenon relaxation time be longer than $T$ [see Figs. 1(b) and (c)], a mixture of water vapour (partial pressure = 6 Pa) and nitrogen (partial pressure = 140 Pa) was introduced into a low-pressure gas cell. The absorption line at 0.557 THz has an expected pressure-broadening linewidth of 10.6 MHz full-width at half-maximum (FWHM) from the self-broadening of water vapour and the collision-broadening induced by nitrogen [20]. Since the absorption relaxation time is determined by the inverse of its linewidth (= 94 ns), the absorption phenomenon relaxes over 23 repetition periods of the THz pulse train. Figure 3(a) shows the absorption spectrum obtained when $\tau = T$. In this experiment, we accumulated 5,000 temporal waveforms for each dFT spectrum, and repeated the spectral interleaving 10-times at a spectral interleaving interval of 25 MHz. We confirmed the absorption spectrum with a linewidth of 25 MHz at 0.557 THz. Although the observed spectral linewidth was limited by the spectral interleaving interval (= 25 MHz) rather than the pressure broadening (= 10.6 MHz), it was 10-times better than the inverse of $\tau$ (= 250 MHz). On the other hand, when $\tau = 0.9995T$, the spectral shape was distorted [see Fig. 3(b); spectral interleaving interval

= 25 MHz]. This spectral distortion results from the partially discontinuous absorption transient due to the failure of the temporal connection. The spectral shape was recovered by introducing null data corresponding to $0.0005T$ restoring the temporal connection [see Fig. 3(c)] although the spectral information may have been somewhat lost by padding null data.

Next, we investigated the achievable spectral resolution limit of the proposed method by measuring the pressure broadening characteristic of the narrow water absorption line. To determine the spectral linewidth, we fitted a Lorentzian function to the measured spectral profile. The red circles in Fig. 3(d) show the FWHM of the observed absorption line with respect to the total pressure of this sample gas, which was varied from 1 Pa to 10 kPa. The green triangles in this graph indicate the spectral interleaving interval. The expected pressure broadening linewidth [20] and room temperature Doppler broadening linewidth [21] for this gas sample are also indicated as blue solid and broken lines in Fig. 3(d). The expected pressure broadening linewidth does not follow a straight line due to the different mixture ratios used for each total pressure. The spectral linewidth determined by THz-dFTS was limited by the spectral interleaving interval or the expected pressure broadened linewidth, whichever is larger. In this way, each datapoint has an infinitesimal spectral resolution, and the spectral interleaving interval finally determines the practical spectral resolution achieved by the proposed method.

2.4 Spectroscopy of low-pressure acetonitrile gas

Finally, we applied THz-dFTS to gas-phase acetonitrile (CH$_3$CN) to demonstrate its capacity to simultaneously probe multiple absorption lines. CH$_3$CN is not only a very abundant species in the interstellar medium but is also a volatile organic gas compounds found in the atmosphere. Since CH$_3$CN is a symmetric top molecule with a rotational constant, *B*, of 9.199 GHz and a centrifugal distortion constant, $D_{JK}$, of 17.74 MHz [22], the frequencies of its rotational transitions are given by

$$v = 2B(J+1) - 2D_{JK}K^2(J+1) \qquad (3)$$

where *J* and *K* are rotational quantum numbers. The first term in Eq. (3) indicates that many groups of absorption lines regularly spaced by 2B (= 18.40 GHz) appear. The second term indicates that each group includes a series of closely spaced absorption lines of decreasing strength. It has been difficult to observe both of these two features simultaneously with conventional THz spectroscopy [23, 24]. Recently, gapless THz-DCS, based on a combination of DCS with spectral interleaving in the THz region, has been successfully used to observe them [15]. However, the huge number of the sampling data produced by this method hinders its extensive use in practical gas analysis.

To confirm the first spectral feature of the symmetric top molecule, we performed broadband THz-dFTS of gas-phase CH$_3$CN at a pressure of 30 Pa. Figure 4(a) shows its absorption spectrum within a frequency range from 0 to 1 THz when the spectral interleaving interval around 0.65 THz was set at 12.5 MHz. 39 groups of absorption lines periodically appeared with a frequency separation of 18.40 GHz. These groups can be assigned to rotational quantum numbers from J = 15 around 0.29 THz to J = 53 around 0.98 THz. The observed dispersion of the peak absorbance is due to the mismatching between the absorption line positions and the spectral sampling positions rather than low SNR. Figure 4(b) shows the magnified spectrum for the 0.6 to 0.7 THz region, indicating that 6 individual groups (J = 32 to 37) have fine spectral features related to the rotational transition of $CH_3CN$ molecules in the ground state and the vibrationally excited state [see red asterisks in Fig. 4(b)].

Next, to confirm the second feature of the symmetric top molecule, we spectrally expanded the absorption line group of J = 34. Nine sharp absorption lines were clearly observed, as shown in Fig. 4(c). By performing multi-peak fitting analysis based on a Lorentzian function, we correctly assigned them to K = 2 to 10. However, two absorption peaks, for K = 0 and 1, were not clear because their frequency separation between them (=15 MHz) is comparable to the spectral interleaving interval (= 12.5 MHz). To spectrally resolve these two absorption peaks, we further reduced the spectral interleaving interval down to 1.25 MHz. This yielded a measured profile containing two fully resolved absorption lines for K = 0 and K = 1, as shown in Fig. 4(d). The spectral linewidth was determined to be 3.8 MHz for K = 0 and 6.1 MHz for K = 1. We also determined their centre frequencies to be 0.64326876 THz for K = 0 and 0.64325750 THz for K = 1. The discrepancy of the centre frequencies from the literature values in the JPL database (see green dashed line) [25] was 1.10 MHz for K

= 0 and 0.12 MHz for K = 1, which were both within the spectral interleaving interval of 1.25 MHz. In this way, we successfully observed both the spectral signatures of $CH_3CN$ characterised by $B$ and $D_{JK}$ using a single instrument.

## 3. Discussion

To highlight dFTS, it is important to compare it with DCS [5-12] because these two methods are analogous. The data volume requirements and achievable spectral resolution of the two techniques are key. We first discuss the total amount of data of the acquired temporal waveform. In DCS, the temporal waveform of multiple periodic signals of a phenomenon is acquired with a time window extending over many repetition periods. The mode-resolved comb spectrum is obtained by calculating its FT, it contains a series of frequency spikes with a frequency spacing equal to the repetition frequency and a linewidth equal to the reciprocal of the time window. For example, if the time window is extended up to $N_p$ times the repetition period, the linewidth is reduced to $1/N_p$ of the frequency spacing, as shown in the upper part of Fig. 5(a). As the resolution and spacing are no longer equal, a frequency gap is created between successive comb modes. Hence the remaining $(N_p-1)/N_p$ frequencies, namely comb gaps, lack any significant information due to the absent of the radiation even though the mode-resolved comb spectrum is composed of a huge number of datapoints. Therefore, the data quantity ratio of the comb modes to the comb gaps, namely, the signal efficiency with respect to all spectral data points, is relatively low. On the other hand, the dFTS spectrum is equivalent to a spectrum in which only the peaks of each comb mode in the upper part of Fig. 5(a) are sampled with infinitesimal spectral resolution, as shown in the upper part of Fig. 5(b). Therefore, all spectral data points contribute to the signal components, and hence the signal efficiency is high. Although a quantitative comparison is given later, the large reduction of the amount of data is a great advantage of dFTS over DCS.

We next discuss the achievable spectral resolution of the two methods. When DCS is combined with the spectral interleaving technique, the spectral resolution will be limited by the comb mode linewidth or the spectral interleaving interval, whichever is larger, as shown in the lower part of Fig. 5(a). On the other hand, in the case of dFTS, the actual resolution is determined by only the spectral interleaving interval, as shown in the lower part of Fig. 5(b). The achievable spectral resolution in the upper of Fig. 5(b) is limited by how long ago the oldest phenomenon in the observation time window is induced. For example, if we set the waiting time for 20 ms before starting the data acquisition, the theoretical limit of the minimum spectral resolution is 50 Hz because all phenomena induced within the waiting time are included in the time window. But in reality, the spectral resolution will be limited by timing jitter between dual lasers because the timing jitter makes the temporal connection difficult due to fluctuation of the temporal magnification factor [26]. The spectral resolution determined by only the spectral interleaving interval is another advantage of dFTS over DCS.

## 4. Conclusions

We demonstrated that the time window size did not limit the spectral resolution in dFTS using precisely periodic radiation only when the time window size is matched to the repetition period. The combination of dFTS with the spectral interleaving technique enables the narrow low-pressure absorption lines to be assigned within the spectral interleaving interval of 1.25 MHz. Further reduction of the spectral interleaving interval may improve the spectral accuracy as well as the spectral resolution. It should also be emphasized that THz-dFTS significantly reduces the required data volume while achieving a spectroscopic performance equal to or greater than THz-DCS.

Although THz-dFTS was demonstrated based on the ASOPS method in this article, the mechanical time-delay scanning in THz-TDS may be also used for THz-dFTS if the requirement $\tau = T$ can be satisfied precisely. Furthermore, it will be possible to apply dFTS to a variety of FTS techniques using other radiation sources. For example, dFTS can be easily implemented in the experimental setup of DCS at other wavelength regions because of their similar setups. Furthermore, FT spectral analysis, FT-NMR, or FT-MS may be combined with dFTS if a repeating phenomenon can be observed using precisely periodic radiation and measured precisely with a time window size equal to one repetition period. The high flexibility of dFTS allowing it to be applied to various types of FTS will accelerate its practical adoption in the fields of science, industry, and medicine.

## A. Methods

### A1. Theory

The FT spectrum $H(f)$ for the measured temporal waveform $h(t)$ of a phenomenon [see Fig. 1(a)] is given by

$$H(f) = \int_{-\infty}^{\infty} h(t)\exp(-2\pi ift)\,dt$$

$$= \cdots + \int_{-T}^{0} h(t)\exp(-2\pi ift)\,dt + \int_{0}^{T} h(t)\exp(-2\pi ift)\,dt + \int_{T}^{2T} h(t)\exp(-2\pi ift)\,dt + \cdots$$

$$= \cdots + \int_{0}^{T} h(t-T)\exp\{-2\pi if(t-T)\}\,dt + \int_{0}^{T} h(t)\exp(-2\pi ift)\,dt + \int_{0}^{T} h(t+T)\exp\{-2\pi if(t+T)\}\,dt + \cdots$$

$$= \sum_{k=-\infty}^{\infty} \int_{0}^{T} h(t-kT)\exp\{-2\pi if(t-kT)\}\,dt \quad (4),$$

where $k$ is an integer. In this case, the spectral resolution is infinitesimal. On the other hand, when the phenomenon repeats at a time period $T$, as shown in Fig. 1(b), the repeating phenomenon $g(t)$ is given by

$$g(t) = \sum_{k=-\infty}^{\infty} h(t-kT) \quad (5).$$

If we observe $g(t)$ for a time window of $0<t<T$, as shown in Fig. 1(c), its FT spectrum $G(f)$ is given by

$$G(f) = \int_{0}^{T} g(t)\exp(-2\pi ift)\,dt$$
$$= \sum_{k=-\infty}^{\infty} \int_{0}^{T} h(t-kT)\exp(-2\pi ift + 2\pi ink)\,dt \quad (6),$$

where $n$ is an integer, and exp(-2πift) = exp(-2πift + 2πink) [see Figs. 1(d) and (e)].

Therefore, if $f_n = n/T$, the corresponding FT spectrum $G(f_n)$ is given by

$$G(f_n) = \sum_{k=-\infty}^{\infty} \int_{0}^{T} h(t-kT)\exp\{-2\pi if_n(t-kT)\}\,dt = H(f_n) \quad (7),$$

indicating that Eq. (7) is equivalent to Eq. (4). That is to say, if a phenomenon repeats with a time period $T$ and we observe it for a finite time window of $0<t<T$, its FT spectrum provides spectral information with an infinitesimal resolution at a frequency $f_n$ (= $n/T$). In other words, infinitesimal resolution can be achieved at a discrete frequency $f_n$ in FTS with a finite time window.

The temporal signal is sampled, and its FT transform is calculated on a computer. The sampled data $g_s(t)$ is expressed by

$$g_s(t) = \sum_{m=0}^{N-1} g(t)\delta\left(t - \frac{mT}{N}\right), \tag{8}$$

where $N$ is the total number of sampling points, and $\delta$ is the delta function. The FT of Eq. (8) is

$$\begin{aligned} G_s(f_n) &= \int_{-\infty}^{\infty} \sum_{m=0}^{N-1} g(t)\delta\left(t - \frac{mT}{N}\right)\exp(-2\pi f_n t)dt \\ &= \sum_{m=0}^{N-1} g(t_m)\exp(-2\pi f_n t_m) \end{aligned} \tag{9}$$

where $t_m = mT/N$. Eq. (9) is the simple dFT of $g_s(t)$ with time window $T$. It is known that information is not lost by sampling when the sampling frequency is faster than two-times the maximum frequency of the data (sampling theorem) [27]. Therefore, discrete spectral data with an infinitesimal spectral resolution is given by taking the simple dFT of $g_s(t)$ sampled with a finite time window

A2. Experimental setup

We used dual mode-locked Er-doped fibre lasers (ASOPS TWIN 250 with P250, Menlo Systems; centre wavelength $\lambda_c$ = 1550 nm, pulse duration $\Delta\tau$ = 50 fs, mean power $P_{mean}$ = 500 mW) for THz-dFTS. Their repetition frequencies ($f_{rep1}$ = 250,000,000 Hz, and $f_{rep2}$ = $f_{rep1}$+$f_{offset}$ = 250,000,050 Hz) and the frequency offset between them ($f_{offset}$ = $f_{rep2}$–$f_{rep1}$ = 50 Hz) were stabilized by using two independent laser control systems referenced to a rubidium frequency standard (Rb-FS, accuracy = 5×10$^{-11}$, instability = 2×10$^{-11}$ at 1 s), as shown in Fig. 2. Furthermore, $f_{rep1}$ and $f_{rep2}$ could be respectively tuned over a frequency range of ±0.8 % by changing the cavity length with a stepper motor and a piezoelectric actuator. After wavelength conversion of the two laser beams by second-harmonic-generation (SHG) crystals, pulsed THz radiation was emitted from a dipole-shaped, low-temperature-grown (LTG), GaAs photoconductive antenna (PCA1) triggered by pump light ($\lambda_c$ = 775 nm, $\Delta\tau$ = 80 fs, $P_{mean}$ = 19 mW), passed through a low-pressure gas cell (length = 500 mm, diameter = 40 mm), and was then detected by another dipole-shaped LTG GaAs photoconductive antenna (PCA2) triggered by probe light ($\lambda_c$ = 775 nm, $\Delta\tau$ = 80 fs, $P_{mean}$ = 9 mW). The optical path in which the THz beam propagated, except for the part in the gas cell, was purged with dry nitrogen gas to avoid absorption by atmospheric moisture. A small portion of the output light from the two lasers was fed into a sum-frequency-generation cross-correlator (SFG-XC). The resulting SFG signal was used to generate a time origin signal for the ASOPS measurement. After amplification with a current preamplifier (AMP, bandwidth = 1 MHz, gain = 4×10$^6$ V/A), the temporal waveform of the output current from PCA2 was acquired with a digitizer (sampling rate = 2×10$^6$ samples/s, resolution = 20 bit) by using the SFG-XC's output as a trigger signal and the frequency standard's output as a clock signal. Then, the time scale of the observed signal was multiplied by a temporal magnification factor of $f_{rep1}/f_{offset}$ (= 250,000,000/50 = 5,000,000) [17]. This sampling rate and this temporal magnification factor enabled us to measure the temporal waveform of the THz electric field signal at a sampling interval of 100 fs during one repetition period. It is important to note that the pulse timing ($f_{rep1}$ and $f_{rep2}$), the trigger timing ($f_{offset}$), and the data acquisition timing in the digitizer were completely synchronized by use of Rb-FS for a common time base. Such excellent synchronization is a critical factor in correctly achieving the temporal connection of many signal pieces at different timings in the temporal waveform.

The spectral interleaving in the dFTS spectrum was performed by incremental increases of $T$, or $f_{rep1}$, with the laser control systems. For example, in the rotational-transition absorption spectroscopy of water, shown in Fig. 3(a) and (b), incremental increases of $f_{rep1}$ and $f_{rep2}$ by 11,220.8 Hz were repeated 10 times while keeping $f_{offset}$ at 50 Hz. In this case, a single shift of $f_{rep1}$ by 0.004488 % resulted in sweeping of the spectral sampling data around the water absorption line at 0.557 THz by 10 % of frequency interval in the dFTS spectrum (= 25 MHz) because the shift of $f_{rep1}$ is multiplied by $n$ (= 2,228 at 0.557 THz) in Eq. (2).

A3. Comparison between THz-dFTS and THz-DCS

We compared the total amount of data of the acquired temporal waveform between THz-dFTS and THz-DCS. For example, in the experiment of Fig. 4(d), we repeated the spectral interleaving 20 times at an interval of 1.25 MHz to obtain the absorbance spectrum. To do so, we acquired 20 temporal waveforms at 20 different $T$ values (sampling interval = 100 fs, time window size = 4 ns), and hence the total number of sampling data was 800,000 [= 40,000 (data/waveform) * 20 (waveforms)]. On the other hand, to perform the same experiment using a combination of THz-DCS with the spectral interleaving [15], one has to first acquire the temporal waveform of the THz pulse train with a time window size equal to 200 repetition periods (= 800 ns) at a sampling step of 100 fs to reduce the comb mode linewidth to 1.25 MHz. Then, one has to acquire 20 temporal waveforms at different $T$ values for 20-times spectral interleaving. In this case, the total number of sampling data is 160,000,000 [= 8,000,000 (data/waveform) * 20 (waveforms)], which is 200 times larger than that in THz-dFTS.

**Acknowledgements**

This work was supported by Collaborative Research Based on Industrial Demand from the Japan Science and Technology Agency, and Grants-in-Aid for Scientific Research No. 26246031 from the Ministry of Education, Culture, Sports, Science, and Technology of Japan. We also gratefully acknowledge financial support from the Renovation Centre of Instruments for Science Education and Technology at Osaka University, Japan. The authors are grateful to Prof. Kaoru Minoshima of the University of Electro-Communications and Prof. Tetsuo Iwata of Tokushima University, Japan, for fruitful discussions.

**Author Contributions**

T. Y. conceived the idea and wrote the paper. Y. I., Y.-D. H, Y. S., and F. H. built the instrument, performed the experiment, and analysed data. M. H. and S. Y. conceived the idea and contributed to manuscript preparation. T. A. supervised the research.

**Competing Financial Interests statement**

The authors declare no competing financial interests.

**Figure Legends**

Fig. 1. Principle of operation. (a) Fourier transform for a measured temporal waveform *h(t)* of a phenomenon (b) Temporal overlapping of multiple repeating phenomena using precisely periodic radiation with a repetition period *T*. (c) Signal acquisition of portions of *h(t)* with different timings with a time window size $\tau$. (d) Fourier transform of portions of *h(t)* with different timings. (e) Temporal connection of portions of *h(t)* with different timings and Fourier transform of the temporally connected *h(t)* without limitation of the time window size. (f) Discrete Fourier transform spectrum for the temporally connected *h(t)* and the spectral interleaving by changing *T*.

Fig. 2. Experimental setup. Rb-FS, rubidium frequency standard; SFG-XC, sum-frequency-generation cross-correlator; SHG, second-harmonic-generation crystals; L, lenses; EDFA, erbium-doped fibre amplifier; OSC, erbium-doped fibre oscillator; PCA1, dipole-shaped low-temperature-GaAs photoconductive antenna for THz emitter; PCA2, dipole-shaped low-temperature-GaAs photoconductive antenna for THz detector; Si-L, silicon lenses; AMP, current preamplifier.

Fig. 3. Absorbance spectrum at the water rotational-transition ($1_{10} \leftarrow 1_{01}$) absorption line with the expected pressure broadening linewidth of 10.6 MHz for (a) $\tau = T$, (b) $\tau = 0.9995T$, and (c) $\tau = T$ achieved by connecting the temporal data for $0.9995T$ and the null data for 0.0005*T*. The spectral interleaving interval is 25 MHz.

Fig. 4. Absorbance spectrum of low-pressure $CH_3CN$ gas within a frequency range (a) from 0.2 to 1 THz, (b) from 0.6 to 0.7 THz, and (c) around 0.6428 THz with a spectral interleaving interval of 25 MHz. Red asterisks in Fig. 4(b) shows the spectral features caused by the rotational transition of $CH_3CN$ molecules in the vibrationally excited state. (d) Expanded absorbance spectrum for two adjacent absorption peaks around 0.643265 THz with a spectral interleaving interval of 1.25 MHz.

Fig. 5. Comparison of the spectral behavior between (a) DCS and (b) dFTS with and without the spectral interleaving.

Reference

[1] Griffiths P. R. & Haseth J. A. D. *Fourier transform infrared spectrometry* (Wiley-Interscience, New York, 2007).

[2] Mittleman D. M. *Sensing with THz radiation* (Springer-Verlag, Berlin, 2003).

[3] Farrar T. C. & Becker E. D. *Pulse and Fourier transform NMR: introduction to theory and methods* (Academic Press, New York, 1971).

[4] Kruppa G. H. & Cassady C. J. *Fourier transform mass spectrometry* (CRC Press, Boca Raton, 2006).

[5] Schiller, S. Spectrometry with frequency combs. *Opt. Lett.* **27**, 766-768 (2002).

[6] Keilmann, F., Gohle, C. & Holzwarth, R. Time-domain mid-infrared frequency-comb spectrometer. *Opt. Lett.* **29**, 1542-1544 (2004).

[7] Yasui, T. *et al.* Terahertz frequency comb by multifrequency-heterodyning photoconductive detection for high-accuracy, high-resolution terahertz spectroscopy. *Appl. Phys. Lett.* **88**, 241104 (2006).

[8] Bernhardt, B. *et al.* Cavity-enhanced dual-comb spectroscopy. *Nature Photon.* **4**, 55-57 (2010).

[9] Yasui, T. *et al.* Fiber-based, hybrid terahertz spectrometer using dual fibre combs. *Opt. Lett.* **35**, 1689-1691 (2010).

[10] Coddington, I., Swann, W. C. & Newbury, N. R. Time-domain spectroscopy of molecular free-induction decay in the infrared. *Opt. Lett.* **35**, 1395-1397 (2010).

[11] Ideguchi, T. *et al.* Adaptive dual-comb spectroscopy in the green region. *Opt. Lett.*

**37**, 4847-4849 (2012).

[12] Hsieh, Y.-D. *et al.* Terahertz comb spectroscopy traceable to microwave frequency standard. *IEEE Trans. Terahertz Sci. Tech.* **3**, 322-330 (2013).

[13] Jacquet, P. *et al.* Frequency comb Fourier transform spectroscopy with kHz optical resolution. *Fourier Transform Spectroscopy 2009* (Vancouver, 2009), FMB2.

[14] Baumann, E. *et al.* Spectroscopy of the methane $\nu_3$ band with an accurate midinfrared coherent dual-comb spectrometer. *Phys. Rev. A* **84**, 062513 (2011).

[15] Hsieh, Y.-D. *et al.* Spectrally interleaved, comb-mode-resolved spectroscopy using swept dual terahertz combs. *Sci. Reports* **4**, 3816 (2014).

[16] Janke, C. *et al.* Asynchronous optical sampling for high-speed characterization of integrated resonant terahertz sensors. *Opt. Lett.* **30**, 1405-1407 (2005).

[17] Yasui, T., Saneyoshi, E. & Araki, T. Asynchronous optical sampling terahertz time-domain spectroscopy for ultrahigh spectral resolution and rapid data acquisition. *Appl. Phys. Lett.* **87**, 061101 (2005).

[18] Klatt, G. *et al.* Rapid-scanning terahertz precision spectrometer with more than 6 THz spectral coverage. *Opt. Express* **17**, 22847-22854 (2009).

[19] Yasui, T. *et al.* Enhancement of spectral resolution and accuracy in asynchronous-optical-sampling terahertz time-domain spectroscopy for low-pressure gas-phase analysis. *Opt. Express* **20**, 15071–15078 (2012).

[20] Seta, T. *et al.* Pressure broadening coefficients of the water vapour lines at 556.936 and 752.033 GHz. *J. Quantum Spectrosc. Radiat. Transfer* **109**, 144–150

(2008).

[21] Sigrist, M. W. *Air monitoring by spectroscopic techniques* (John Wiley & Sons, New York, 1994).

[22] Pearson J. C. & Müller H. S. P. The submillimeter wave spectrum of isotopic methyl cyanide. *Astrophys. J.* **471**, 1067-1072 (1996).

[23] Mittleman, D. M. *et al.* Gas sensing using terahertz time-domain spectroscopy. *Appl. Phys. B* **67**, 379–390 (1998).

[24] Matsuura, S. *et al.* High-resolution terahertz spectroscopy by a compact radiation source based on photomixing with diode lasers in a photoconductive antenna. *J. Mol. Spectrosc.* **187**, 97–101 (1998).

[25] Rothman, L. S. *et al.* Submillimeter, millimeter, and microwave spectral line catalog. *J. Quant. Spectrosc. Radiat. Transf.* **60**, 883–890 (1998).

[26] Bartels, A. *et al.* Ultrafast time-domain spectroscopy based on high-speed asynchronous optical sampling. *Rev. Sci. Instrum.* **78**, 035107 (2007).

[27] Bracewell R. N. *Sampling and series: Fourier transform and its applications Ch. 10* (MacGraw-Hill, Cambridge, 1999)

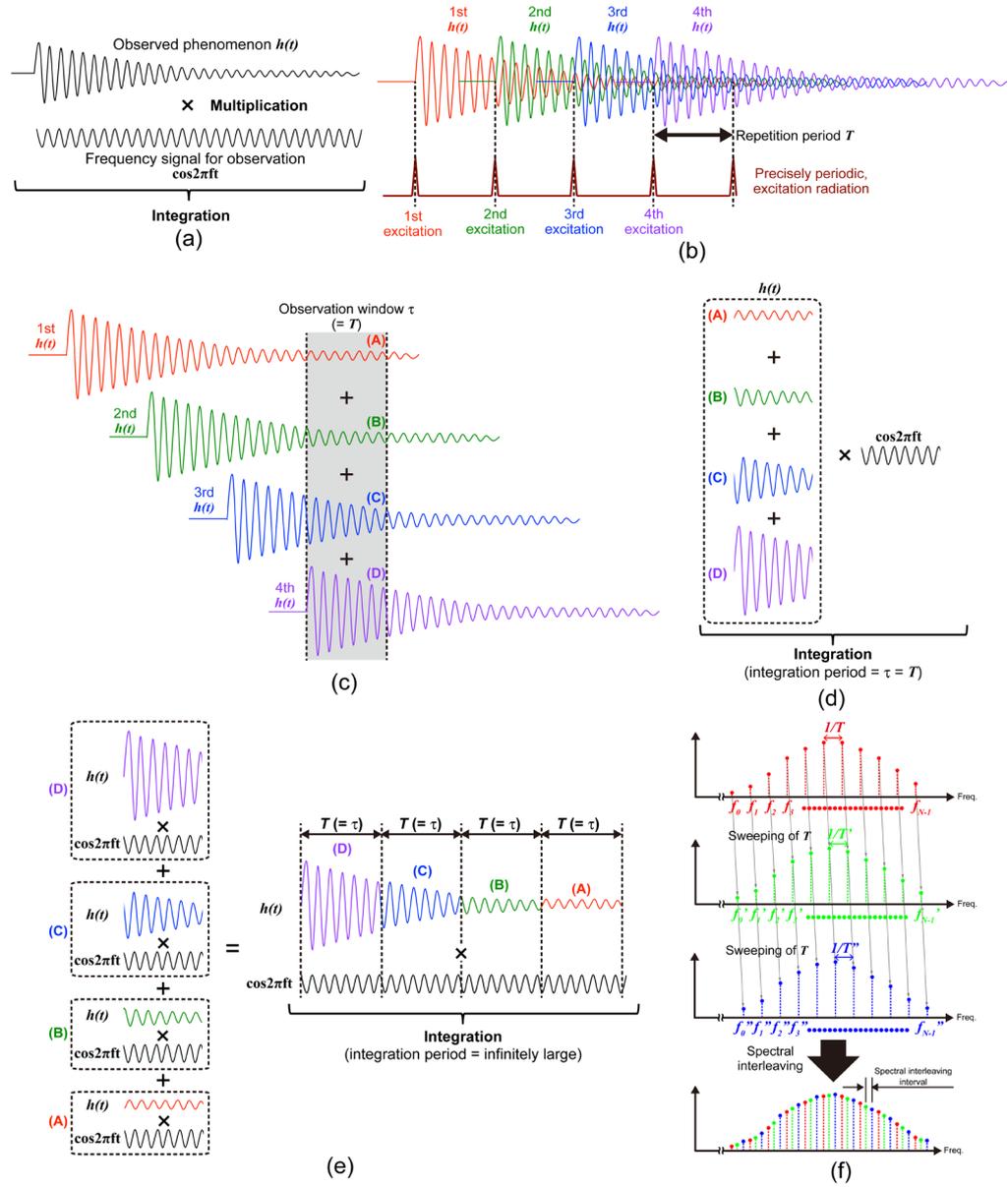

Fig. 1

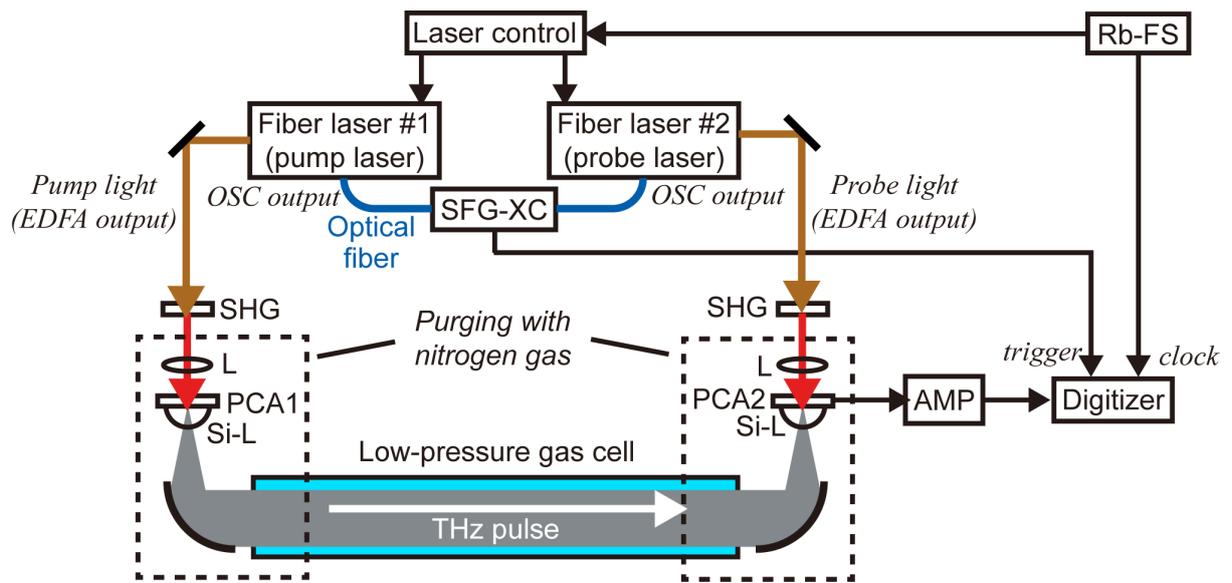

Fig. 2

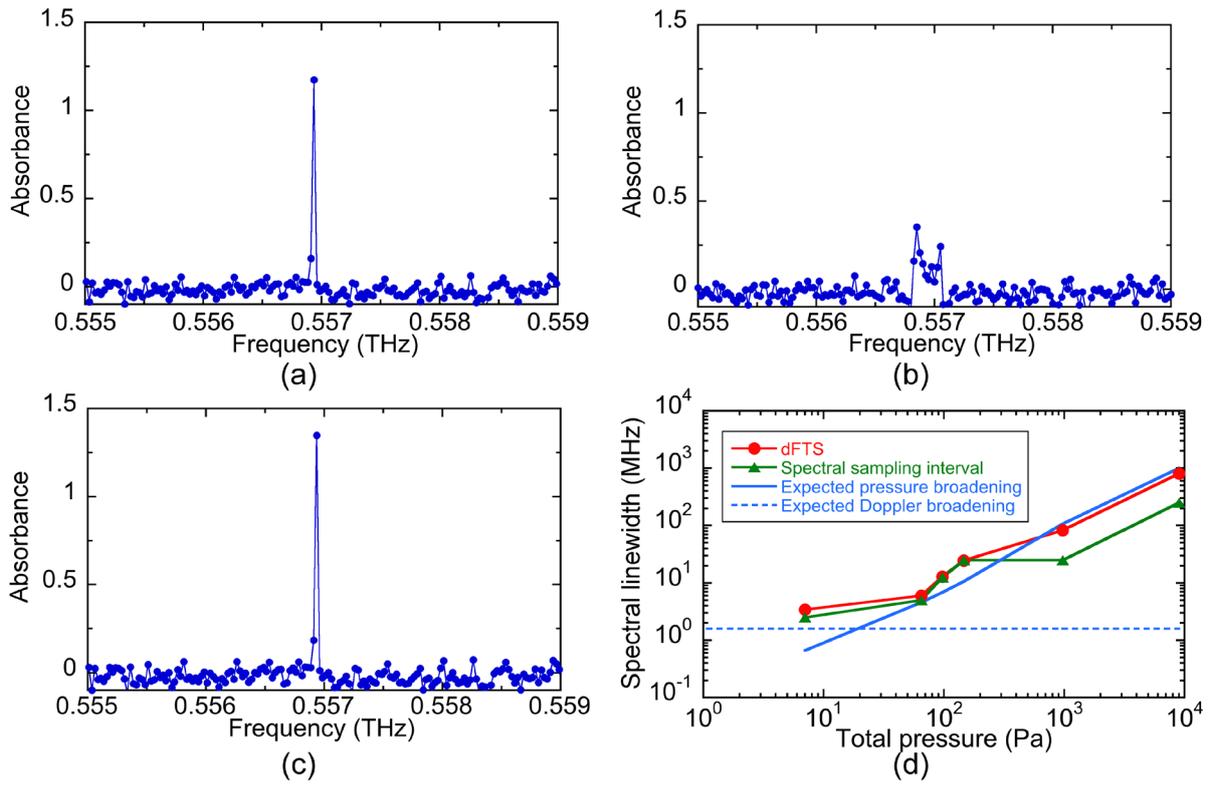

Fig. 3.

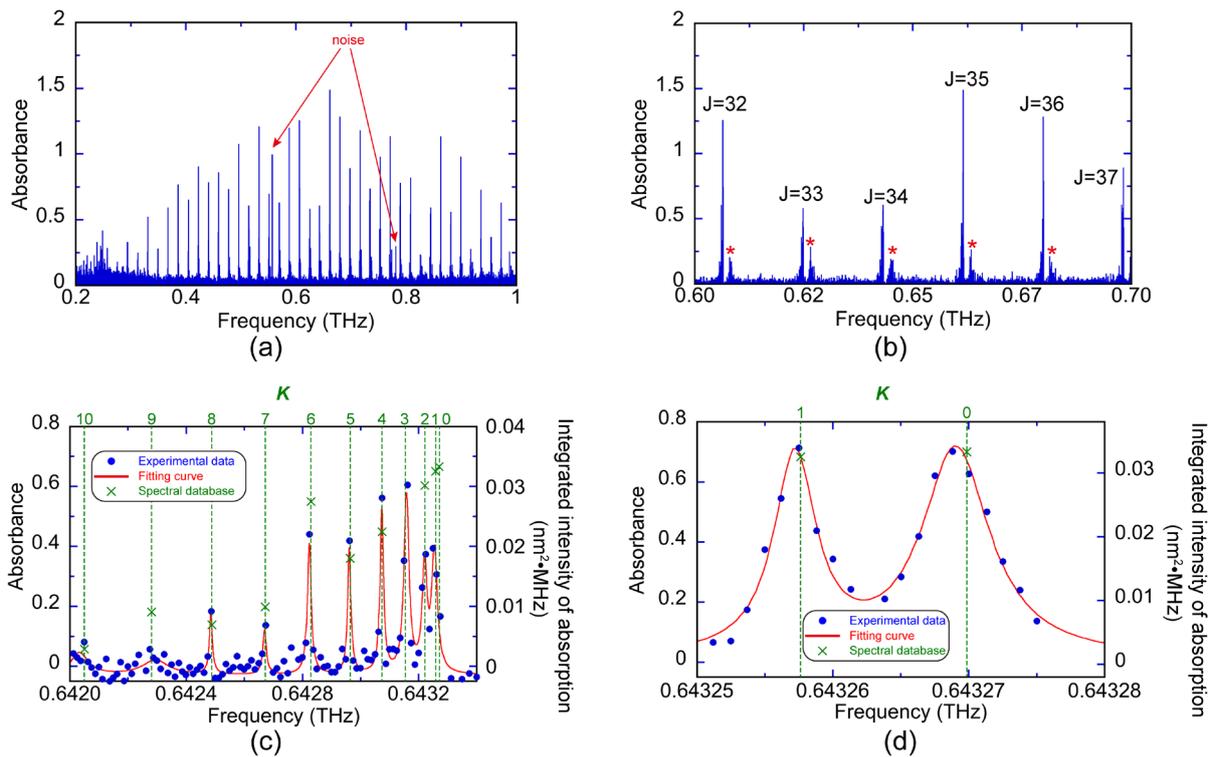

Fig. 4.

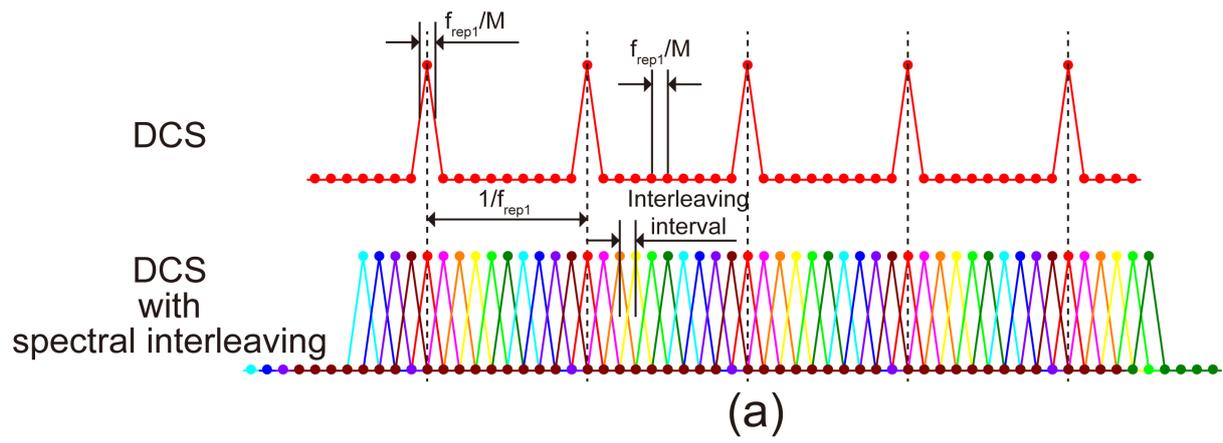

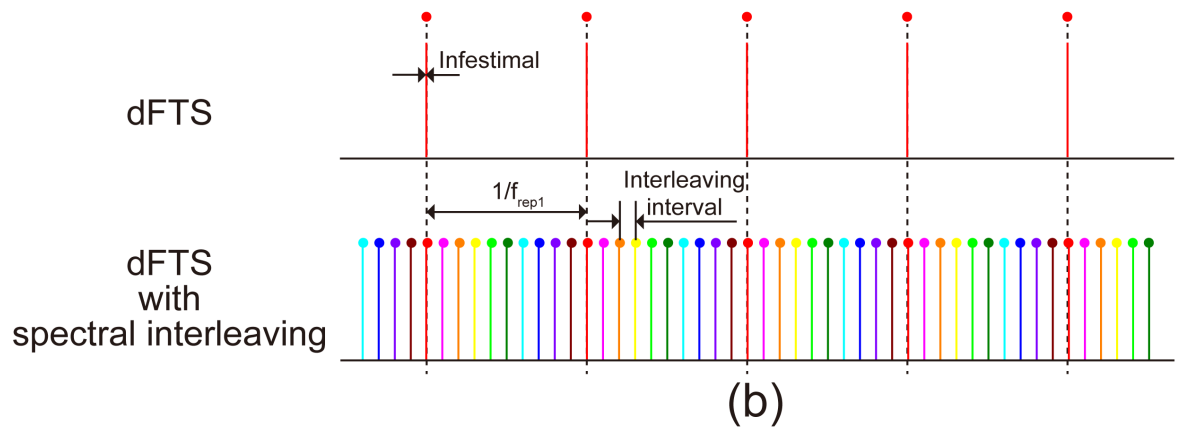

Fig. 5.